\documentstyle[epsf,pra,aps]{revtex}

\begin{document}
\draft

\def\del{\partial}

\title{Theory of suppressed shot-noise at $\nu=2/(2p+\chi)$}
\author{K.-I. Imura$^1$ and K. Nomura$^2$}
\address{$^1$ Department of Applied Physics, University of Tokyo,
Hongo 7-3-1, Tokyo 113-8656, Japan}
\address{$^2$ Department of Basic Science, University of Tokyo,
Komaba 3-8-1, Tokyo 153-8902, Japan}
\date{\today}
\maketitle

\begin{abstract}
We study the edge states of fractional quantum Hall liquid at
bulk filling factor $\nu=2/(2p+\chi)$ with $p$ being an even integer and
$\chi=\pm 1$.
We describe the transition from a conductance plateau $G=\nu G_0=\nu e^2/h$
to another plateau $G=G_0/(p+\chi)$ in terms of chiral Tomonaga-Luttinger
liquid theory.
It is found that the fractional charge $q$ which appears in the classical
shot-noise
formula $S_{I}=2q \langle I_b \rangle$ is $q=e/(2p+\chi)$ on the conductance
plateau at
$G=\nu G_0$ whereas on the plateau at $G=G_0/(p+\chi)$ it is given by
$q=e/(p+\chi)$.
For $p=2$ and $\chi=-1$ an alternative hierarchy constructions is also
discussed to explain the suppressed shot-noise experiment at bulk filling factor
$\nu=2/3$.
\end{abstract}



The fractional quantum Hall (FQH) effect is a phenomenon observed in
a two-dimensional electron system subjected to a strong perpendicular
magnetic field.
Due to the interplay between the strong magnetic field and interaction
among the electrons as well as weak disorder, the transverse resistivity
shows a plateau behavior. \cite{tsui}
For a filling factor $\nu=1/$(odd integer), the theory predicts fractionally
charged quasiparticles with charge $\nu e$. \cite{bob}
Rececnt shot noise experiments in a two-terminal FQH system with a
point-like
constriction seem to be consistent with this theoretical
prediction.\cite{cglat}
Another aspect of edge tunneling experiments is non-linear scaling behaviors
of the $I-V$ characteristics. \cite{chang,grayson}.
The chiral Tomonaga-Luttinger (TL) liquid theory
seems to be successful at least as an effective theory to describe
such non-linear behaviors of the tunneling current.
\cite{wen1}

Let us consider a two-terminal Hall bar geometry
where the bulk FQH liquids has both upper and lower edges.
Applying the negative gate voltage to squeeze the Hall bar,
one can introduce a depleted region of electrons.
This structure, called a point contact (PC),
introduces the back-scatterred current $I_b$ due to the
quasiparticle tunnneling through the bulk FQH liquid.
In the TL model backward scattering plays the role of periodic
potential barrier with respect to the bosonic fields.
\cite{KFFN}
Based on this model subsequent theoretical works
have shown that the fractiopnal charge $q$ of a quasiparticle could in
principle
be detected in terms of the quantum shot noise
in the weak back-scattering phase,\cite{chmn}
\begin{equation}
S_{I_b}(\omega)=\int_{-\infty}^{\infty}dt\cos\omega t
\langle\{I_b(t),I_b(0)\}\rangle,
\end{equation}
where $I_b=\nu e^2/h -I$.
The shot noise spectrum reduces to the classical shot noise formula
$S_{I_b}=2q \langle I_b\rangle$ for $\omega=0$ (white noise).

The bulk hierarchy theory for filling factors $\nu=m/(mp+\chi)$
predicts the existence of fractionally charged quasiparticle with
charge $q=e/(mp+\chi)$, which differs from $\nu e$ for $m\neq 1$.
\cite{jain}
In fact Reznikov et. al. observed a charge $e/5$ in the shot-noise
experiment at $\nu=2/5$, \cite{rez} which seems enough to convince us
that the fractional charge observed in the shot-noise experiment is
neither a filling factor nor a conductance,
but the charge of current-carrying quasiparticles.

In this paper we describe the transition from a conductance plateau $G=\nu G_0=\nu e^2/h$
to another plateau $G=G_0/(p+\chi)$ in terms of
the reconstruction of two-channel edge modes.
For zero gate voltage a two-channel edge mode is derived from the bulk hierarchy
theory at $\nu=2/(2p+\chi)$. \cite{wen2}
A finite gate voltage introduces the tunneling of quasiparticles
through the bulk FQH liquid. These tunnelings play the role of scattering potential
barrier for the $1+1$-dimensional effective theory.
The tunneling amplitude, or equivalently the height of potential barrier for
each channel is in general quite different.
As the gate voltage is increased, one of the phases corresponding to the 
stronger tunneling channel tends to be pinned but the other remains almost free.
We identify this situation as the plateau of conductance at $G=G_0/(p+\chi)$.
The shot noise spectrum is calculated perturbatively in these plateau region.
The results are compared with the recent experiment of the suppressed shot-noise measurement.

\vspace{0.1cm}
We describe the edge states of FQH liquid at $\nu=2/(2p+\chi)$
by the following two-component chiral TL liquid theory.
\cite{wen2,mc}
\begin{equation}
{\cal L}_{\rm TL}=
-{1\over 4\pi}K^{\alpha\beta}
{\del\phi_\alpha^+\over\del t}
{\del\phi_\beta^-\over\del x}
-{1\over8\pi}V^{\alpha\beta}
\left(
\frac{\partial\phi_\alpha^+}{\partial x}
\frac{\partial\phi_\beta^+}{\partial x}
+
\frac{\partial\phi_\alpha^-}{\partial x}
\frac{\partial\phi_\beta^-}{\partial x}
\right)
\end{equation}
where $\phi^\pm=\phi^u\pm\phi^l$ with $\phi^u (\phi^l)$ being the edge mode
propagating near the upper (lower) boundary of the system.
$\alpha, \beta=1,2$ corresponds to the Landau level indices for the
composite
fermions (pseudospin) or the real spin indices.
Another basis on which we will work is the charge-(pseudo)spin basis,
$\phi_{\rho,\sigma}=\phi_1\pm\phi_2$.

Universal properties of bulk FQH liquid could be classified by
the so-called $K$-matrix and charge vector $t$.
The standard construction of the $K$-matrix at a hierarchical
filling factor $\nu=m/(mp+\chi)$ ($m$: integer, $p$: even integer,
$\chi=\pm 1$ gives us $K=\chi I_m +pC_m$ in the unitary basis $t^T=(1,1)$,
\cite{zee}
where $I_m,C_m$ are $m\times m$ identity and pseudo-identity matrices.
For $m=2$ it reduces to
\begin{equation}
K= \left[
\begin{array}{cc}
p+\chi & p \\
p & p+\chi
\end{array}
\right],
\end{equation}
which can be identified with the matrix $K$ in Eq.(1).
After a linear transformation we obtain a charge boson and a neutral
boson, which has
$K_\rho=\nu/2=1/(2p+\chi), K_\sigma=1$,
respectively.
$\chi$ correspons to the chirality of the edge modes.
For $\chi=1$ charge and (pseudo)spin modes propagate in the same
direction (co-propagate),
whereas for $\chi=-1$ they are counter-propagating.
In this paper we focus on the (bulk) filling factors $\nu=2/(2p+\chi)$
with $p$ being an even integer and $\chi=\pm1$.

Let us consider $\chi=1$, e.g., $\nu=2/5$.
The elementary quasiparticles are predicted to have charge $e/(2p+1)$,
and not $2e/(2p+1)$. They can be written in terms of the boson fields
near the edge of the sample as,
$e^{i\phi_\rho^{u,l}/2}e^{\pm i\phi_\sigma^{u,l}/2}$.
For $p=2$ we identify these elementary quasiparticles with charge $e/5$
as the current-carrying particles in the recent shot-noise experiment
at $\nu=2/5$.
Due to the finite amount of energy to create a quasiparticle,
the tunneling of quasiparticle is allowed only in the vicinity
of PC, which induces the following potential barrier,
\begin{equation}
{\cal L}_{\rm tun}=
u_1\delta (x)\cos\phi_1^{+}+
u_2\delta (x)\cos\phi_2^{+}.
\end{equation}
Here we assumed that the tunneling amplitude 
of quasiparticles with charge $2e/5$ is negligiblly
small.\cite{cmnt}
In Ref. \cite{rez} the two-terminal conductance shows
a transition from a plateau at $G=(2/5)e^2/h$ to another plateau at
$G=(1/3)e^2/h$ as the constriction is increased.
On this plateau they observed a current-carrying particle with charge
$e/3$. The filling factor near the PC is considered to be $\nu=1/3$.
Hence a single-channel edge mode with $K_\rho=1/3$ is expected near
the PC. (Fig.1)

\begin{figure}[h]
\input epsf
\epsfbox{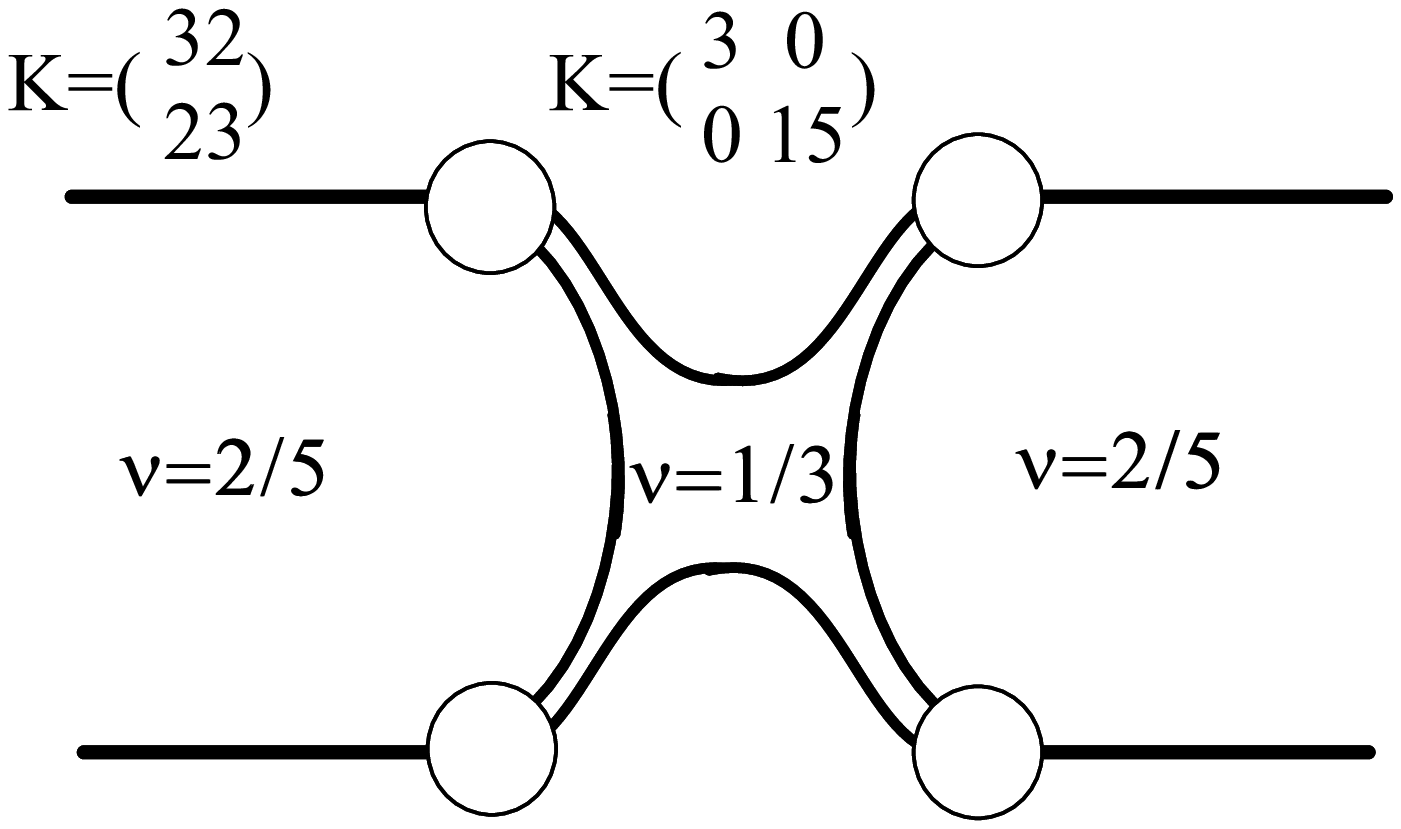}
\vspace{0.1cm}
\caption{Edge reconstruction for $\nu=2/5$:
The two-terminal conductance shows a transition from
a plateau at $G=(2/5)e^2/h$ to another plateau at
$G=(1/3)e^2/h$ as the constriction is increased.
On the second plateau the filling factor near the PC would be $\nu=1/3$.
A single-channel edge mode with $K_\rho=1/3$ is expected near the PC.
This means that there is a step of filling factors by $\Delta\nu=1/15$
between the bulk and PC. Another edge moge is expected along this step,
which has only one channel with $K_\rho=1/15$.
Whereas near the PC a single-channel edge mode with
$K_\rho=1/3$ is expected, say, near the upper boundary of
the sample. These two edge modes approach one another as they propagate
far from the PC. In this region we can describe the edge modes by
an effective action (2) with $K={\rm diag}(3,15)$.
The quasiparticles corresponding to the two independent
FQH liquids have charges $e/3$ and $e/15$, respectively.
However the quasiparticles with charge $e/15$ are not detected
in the shot-noise experiment since the tunneling current is not affected by
such quasiparticles.
On the other hand in the region where the physics is completely unaffected
by the gate, the matrix $K$ in (2) should be given by Eq. (3) with
$p=2,\chi=1$,
since the elementary quasiparticle should have charge $e/5$.
Hence an ''edge reconstruction'' is expected from $K={\rm diag}(3,15)$ to it
as they propagate from the vicinity of PC to the edge of ''bulk'' FQHL.
}
\end{figure}

For $\nu=2/3$ the situation is slightly different.
We have at least two constructions of the edge modes for this filling
factor. \cite{mc}
One comes from the standard hierarchy in the bulk
(3) with $p=2,\chi=-1$.
The elementary quasiparticles have a charge $e/3$,
which is consistent with the shot-noise
experiment at $\nu=2/3$. \cite{cglat}
The scattering potential barrier due to the tunneling of quasiparticles at
the PC
takes the same form as Eq. (3).
However a naive shot-noise experiment cannot distinguish between this
standard hierarchy construction and the second construction,
which is called two independent $\nu=1/3$ Laughlin states,
$K= 3 I_2$.\cite{mc}
The elementary quasiparticles have a charge $e/3$ again.
In order to distinguish the above two states in terms of the suppressed
shot-noise measurement we have to study the plateau transition
as the gate voltage is increased, to which we will come later.

\vspace{0.1cm}
Let us begin with a general discussion on the
plateau transition at bulk filling factor $\nu=2/(2p+\chi)$
in terms of our TL model: (2)+(4) with (3).
Consider the renormalization group (RG) phase diagram in the
$u$-plane, where $u =(u_1,u_2)$.
The direction of flow is defined such that it scales to lower energies.
We have at least four fixed points in our $u$-plane;
$u=(0,0), (0,\infty), (\infty,0)$ and $(\infty,\infty)$.
The origin of this plane
corresponds to the conductance plateau at $G=\nu e^2/h$.
A standard RG analysis shows that
only the fixed point at $u=(\infty,\infty)$, which we indentified
as a Hall insulator,
is infra-red (IR) stable, since both $u_1$ and $u_2$ are relevant
near these four fixed points except for $p=2,\chi=-1$.
We introduce a small negative gate voltage to the system
on the conductance plateau at $\nu=2/(2p+\chi)$.
Let us start our RG flow from the point where $u_1\ll u_2\ll qV$,
which is in the vicinity of IR-unstable fixed point $u=(0,0)$.
In our RG analysis $qV$ plays a role of high-frequency cut-off of the system.
The assumption that the tunneling amplitudes are different for
different species of quasiparticles seems to be quite natural in the
experiment.
\cite{de-p2}
In the following we fix the gate voltage and consider the zero temperature
for simplicity. Then the scaling is controlled by the voltage difference
between the two reservoir, or $qV$.
As the voltage is decreased, both $u_1$ and $u_2$ scale to
larger values. But due to the assumption $u_1\ll u_2\ll 1$,
$u_2$ increases much faster than $u_1$. Hence
the RG flow passes near the point $u=(0,\infty)$.
One finds that $u=(0,\infty)$ is a saddle point of
the RG flow corresponding to the $G=G_0/(p+\chi)$ conductance plateau.
After passing through this region, our RG flow goes to the attractive
fixed point $u=(\infty,\infty)$, which corresponds to the zero
conductance except for $p+\chi=1$.
Remember that the filling factor $\nu=2/3$ corresponds to this
exceptional case.
In the following the shot-noise spectrum is calculated pertubatively 
for (1) small $u_1,u_2$ and (2) small $u_1$ with $u_2\rightarrow\infty$.

\vspace{0.1cm}
\noindent
(1) Shot-noise spectrum for small $u_1,u_2$:
In the vicinity of our starting point of the RG flow,
where $u_1,u_2\ll qV$, we can employ a perturbation
theory with respect to $u_1$ and $u_2$.
Let us begin with expressing the tunneling current operator
$I_b$ in terms of the bosonic field,
$I_b(t)=iq(u_1 e^{i\phi_1^{+}(x=0)}+u_2 e^{i\phi_2^{+}(x=0)})+{\rm
H.c.}$,
where $q=(\nu/2)e=e/(2p+\chi)$.
The voltage difference between the two resevoirs can be introduced
by letting $u_{1,2}\rightarrow u_{1,2}e^{-iqVt}$.\cite{mahan}
The current-current correlation is calculated to be
\begin{equation}
\langle I_b(t) I_b(0) \rangle
=2 q^2(|u_1|^2+|u_2|^2) \cos (qVt)
\left({a^2\over a^2+v_c^2 t^2}\right)^{\nu/2^2}
\left({a^2\over a^2+v_s^2 t^2}\right)^{2/2^2},
\end{equation}
where we have introduced a short distance cutoff $a$.
$v_c,v_s$ are diagonal components of the matrix $V$ in the charge-spin
basis.
Eq. (5) gives us the noise spectrum up to the order of $|u|^2$,
\begin{equation}
S_{I_b}=q \langle I_b \rangle
\left(
\left|1-{\omega/qV}\right|^{\nu/2}+
\left|1+{\omega/qV}\right|^{\nu/2}
\right),
\end{equation}
where we have used \cite{wen1}
\begin{equation}
\langle I_b \rangle
=\nu {e^2\over h}-\langle I \rangle
={2\pi q\over \Gamma[\nu/2 +1]}
\left(|u_1|^2+|u_2|^2\right)
{a^{\nu/2 +1}\over v_c^{\nu/2}v_s} (qV)^{\nu/2}.
\end{equation}
For $|\omega|\ll qV$ we recover the classical shot-noise
formula with $q=(\nu/2)e=e/(2p+\chi)$.

\vspace{0.1cm}
\noindent
(2) Shot-noise on $G=G_0/(p+\chi)$ plateau:
Let us consider the case where either $u_1$ or $u_2$ becomes
infinitely large, e.g., consider $u=(u_1,\infty), u_1\ll qV$.
In this case the phase $\phi_2(x=0)$ is pinned at one of the minima of
infinitely deep scattering potential $u_2$.
We are allowed to employ a perturbation theory with respect to
$u_1$ with the phase $\phi_2(x=0)$ fixed.
One obtains basically the same result as the single-channel case
with $K_\rho=1/(p+\chi)$;
\begin{eqnarray}
S_{I_b}&=&q \langle I_b \rangle
\left(
\left|1-{\omega/qV}\right|^{{2\over p+\chi}-1}+
\left|1+{\omega/qV}\right|^{{2\over p+\chi}-1}
\right),
\nonumber \\
\langle I \rangle
&=&{1\over p+\chi} {e^2\over h}
-{2\pi q\over \Gamma[2/(p+\chi)]}|u_1|^2
a^{{2\over p+\chi} -2} v_1^{-{2\over p+\chi}} (qV)^{{2\over p+\chi}-1},
\end{eqnarray}
where $q=e/(2p+1)$, and $v_1$ is a renormalized velocity of the first channel.

\vspace{0.1cm}
Let us think of applying Eqs. (5)-(8) to $\chi=-1$.
When $\chi=-1$, the edge modes are counter-propagating,
which might lead us to qualitatively different physics
if random equilibration between the edges should be taken into acount.
\cite{KFP}
But in the following we switch off ramdom impurity scattering
so that the physics for small $u_1, u_2$ is well described
by Eqs. (5)-(7).
For $p=2,\chi=-1$ one sees that the RG phase diagram has two domains,
i.e., a weak scattering phase and a strong scattering phase,
separated by a separatorix. (Fig. 2)
However even in this case Eq.(5)-(8) still make sence in 
the the weak scattering phase of the RG phase diagram.
However as the RG flow goes into the strong scattering phase
the system begins to look differently from the other cases.
In the vicinity of, say, $u=(0,\infty)$, a plateau behavior of the
conductance
is expected at $G=e^2/h$, on which the physics is essentially described
by a single-channel edge mode. The difference is in that this edge mode
is robust against quasiparticle tunnling through the bulk, i.e., $u_1$
is marginal for the ideal case, and becomes irrelevant when the interaction
with
another counter-propagating edge mode with $K_\rho=1/3$ is taken into
accouont.
\cite{prb} (Fig. 2)
This means that we have two additional attractive fixed points in the
present case at $u=(0,\infty)$ and at $u=(\infty,0)$.
A separatorix is expected between the weak and strong coupling phase,
since both $u_1$ and $u_2$ are attractive in the vicinity on
$u=(\infty,\infty)$.
At any rate the conductance shows a transition from the $G=(2/3)e^2/h$
plateau to another plateau $G=e^2/h$ as the costriction is strengthened.
Therefore one way to validate the standard hierearchy state (3) is
to observe a charge $q=e$ in the shot-noise experiment at bulk filling
factor $\nu=2/3$.

\begin{figure}[h]
\input epsf
\epsfbox{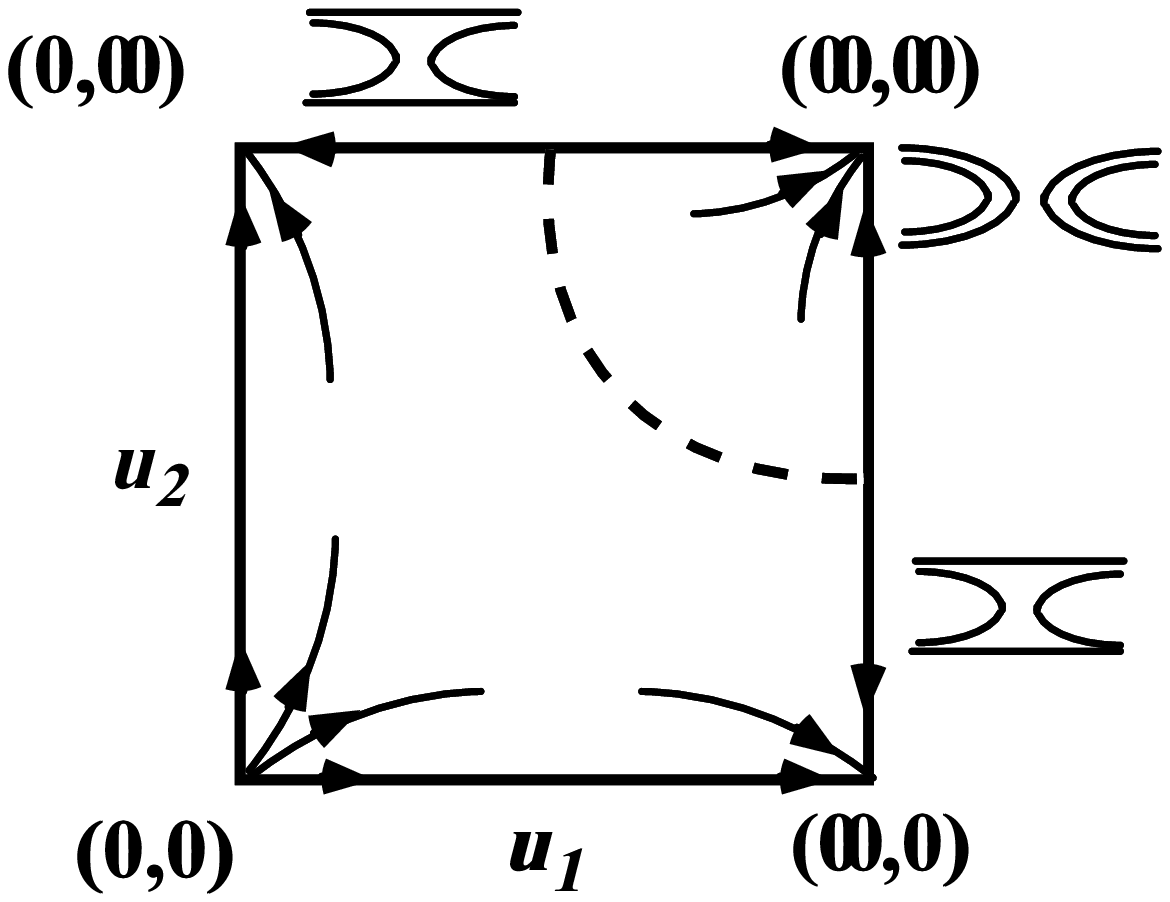}
\vspace{0.1cm}
\caption{RG flow diagram in the $u$-plane for $K=-I_2+2C_2$:
Paths of RG flow are shown for tunneling amplitudes of quasiparticles
$u_1, u_2$. The direction of flow is headed for lower energies.
$u=(0,0)$ corresponds to the conductance plateau at $G=\nu e^2/h$,
but this point is infra-red (IR) unstable in the RG sense.
The RG flow goes toward either of the IR-stable fixed points
$u=(0,\infty)$ or $u=(\infty,0)$ depending on the initial value of
$u_1$ and $u_2$.
They corresponds to an integer quantum Hall (IQH) conductance plateau
$G=e^2/h$.
However we have another attractive (IR-stable)
fixed point at $u=(\infty,\infty)$ corresponding to the zero
conductance.
Between $(\infty,\infty)$ and $(0,\infty)$
or between $(\infty,\infty)$ and $(\infty,0)$
one is supposed to work on $K={\rm diag}(1,-3)$,
since another edge moge is expected along the
step of filling factors $\Delta\nu=-1/3$ between the bulk and PC,
which plays the role of second channel.
The phase diagram of this system
has a separatorix between the two IR-stable fixed points
when inter-channel interactions are taken care of.
Therefore our phase diagram has two domains and the $G=\nu e^2/h$
conductance plateau is not smoothly connected to the
zero conductance phase in a RG sense.
}
\end{figure}

However reports on the suppressed shot-noise measurements at bulk
filling factor $\nu=2/3$ seems to be inconsistent with the standard
hierarchy construction (3). \cite{cglat}
One is inclined to study an alternative construction. Let us choose
$K=3I_2$. The phase diagram is trivial for this construction.
In the weak scattering phase one obtains $q=e/3$. But on the second
plateau $u\sim(0,\infty)$ or $u\sim(\infty,0)$ one also gets
$e/3$. After passing this second plateau at $G=G_0/3$ the conductance
goes smoothly to zero, which seems more realistic for the experiments.

Finally we compare our results with the ``global phase diagram''
in the quantum Hall effect.\cite{KLZ}
We described the plateau transition at bulk filling factor $\nu=2/(2p+\chi)$
in view of the RG flow phase diagram for $1+1$-dimensional effective theory.
As the system scales to lower energies, succeessive transitions
of the conductance are expected: 
$G= 2G_0/(2p+\chi)\rightarrow G_0/(p+\chi)\rightarrow 0$.
The first transition corresponds
to the one from a $\nu=2$ integer quantum Hall (IQH) state to a $\nu=1$ IQH state
for composite fermions, each carrying two flux quanta.
A direct (an anomalous) transition from $G= 2G_0/(2p+\chi)$ to a Hall insulator \cite{mano}
is allowed only when both $u_1$ and $u_2$ are the same order.
Otherwise our RG flow picture is consistent with the one in Ref. \cite{KLZ}
except for $p=2, \chi=-1$.
Another difference is that the direction of the transition is specified in 
our case.
A generalization to the filling factors $\nu=m/(mp+\chi)$
is straightforward, but not discussed here in more detail.

Before ending this paper we would like to comment that the transverse
resistivity $\rho_{xy}$ and the tunneling exponent $\alpha$ behave
quite differently as a function of $1/\nu$.
The plateau behavior of $\rho_{xy}$ is the essence of FQH effect.
Observed factional charges in the shot-noise experiments look also
quantized on the plateaux of the condutance.
However it is not the case with the exponent $\alpha$ for the tunneling into
FQH liquids. The experiment by Grayson et. al. shows that the exponent
$\alpha$ for the $I-V$
characteristics looks continuously varying as $\alpha\sim 1/\nu$
even in a given plateau. \cite{grayson}
Therefore our present goal would be a unified description of the plateau and
continuous behaviors of physical quantities.

In conclusion we studied the transition of conductance from
the bulk value $G=\nu G_0=\nu e^2/h$ to another plateau $G=G_0/(p+\chi)$
in terms of chiral Tomonaga-Luttinger liquid theory.
We calculated the shot-noise spectra near these conductance plateaux
at zero temperature.
We found that the suppressed shot-noise measurement at bulk filling factor
$\nu=2/5$ is well described by the standard construction of $K$-matrix.

\acknowledgements
We are grateful to T. Martin, M. Reznikov, Y. Morita, K.-V. Pham,
D.Yoshioka and N. Nagaosa for useful discussions.
This work was partly supported by COE, Priority Areas Grants and
partly by Grant-in-aid for Scientific Reserch (c) 10640301 
from the Ministry of Education, Science, Sports
and Culture of Japan.
K.-I. I. is also supported by JSPS Research
Fellowships for Young Scientists.

\end{document}